# STUDY THE ISOMERIC RATIOS IN PHOTONUCLEAR EXPERIMENTS ON THE LUE-40 LINAC RDC "ACCELERATOR" NSC KIPT

*I.S. Timchenko[1,2], O.S. Deiev[2], S.M. Olejnik[2], S.M. Potin[2], V.A. Kushnir[2],*
*V.V. Mytrochenko[2,3], S.A. Perezhogin[2]*
[1]*Institute of Physics, Slovak Academy of Sciences, SK-84511 Bratislava, Slovakia;*
[2]*National Science Center "Kharkov Institute of Physics and Technology", Kharkiv, Ukraine;*
[3]*CNRS/IJCLAB, 15 Rue Georges Clemenceau Str., Orsay, France*
*E-mail: timchenko@kipt.kharkov.ua, iryna.timchenko@savba.sk*

The results of photoproduction of isomeric pairs from the ($\gamma$,xn) and ($\gamma$,pxn) reactions on nuclei from Zr up to Ta at the bremsstrahlung end-point energy range 30…100 MeV are systematized. Measurements were performed at the electron beam of the LUE-40 linac RDC "Accelerator" NSC KIPT using the activation method and off-line $\gamma$-ray spectrometric technique. The theoretical isomeric ratios *IR* for the nuclei-products of studied reactions were calculated using the cross-sections $\sigma(E)$ from the TALYS1.95 code for six different level density models. The data from the international databases were additionally used to compare experimental results and theoretical predictions. The obtained results let us make conclusions about trends of values and energy dependence of *IR* in studied photonuclear reactions. Further directions of experimental research in the RDC "Accelerator" NSC KIPT are discussed.

## INTRODUCTION

Photonuclear reactions are accompanied by the emission of a nucleon or a group of nucleons from a compound nucleus. This leads to an excited state of the final nucleus, the discharge of excitation energy which occurs in a time of $10^{-12}…10^{-17}$ sec. However, in some cases, at low energy of the excitation level and a high degree of forbidden transition, long-lived excited states of atomic nuclei occur. Such states are called isomeric or metastable, and their half-lives can vary from nanoseconds to many years [1].

There are nuclei with two unstable states, metastable (m) and ground (g) states. This gives a possibility to determine the population of the metastable state of a given nucleus relative to its ground state, or obtained isomeric ratio *IR*. This value is defined as the ratio of the cross-sections for the nucleus formation in the metastable state to the ground state [2 - 4]. Also in literature, the value of the isomeric ratio is given as the ratio of the cross-sections for the nucleus formation in the high spin state and low spin state. In experiments with bremsstrahlung flux, the isomeric ratio *IR* of the nuclei-products of the studying reaction is measured [4 - 6] as the ratio of flux-averaged cross-sections or reaction yields. This quantity depends on the spin of the target nucleus and the intake angular momentum determined by the mass and the energy of projectile particles.

The study of the isomeric ratios using photonuclear reactions has the advantage that the gamma quantum introduces a small angular momentum and does not change the nucleon composition of the compound nucleus. The data on the isomeric ratios of the reaction products make it possible to investigate questions related to nuclear reactions and nuclear structure: the spin dependence of the nuclear level density, angular momentum transfer, nucleon pairing, and shell effects. It allows to refine the theory of gamma transitions and to test theoretical nuclear models [7 - 11]. The experimental results in the energy range above the giant dipole resonance (GDR) and up to the pion production threshold are of interest because the mechanism of the nuclear reaction changes [12]. As shown in Ref. [13], in this energy range (30…145 MeV) the reaction mechanism changes from the dominant giant dipole resonance to the quasideuteron mechanism.

Currently, investigations are being performed to summarize/systematize the available data on isomeric ratios and compare them with theoretical predictions. For example, in [14] the data on *IR* obtained in reactions with light particles are systematized and compared with the results of calculations in the TALYS code [15].

In [16] the authors measured the isomeric ratios in odd-odd nuclei of high spin isomers $^{196}$Au, $^{182}$Ta and $^{194}$Ir produced in $^{197}$Au($\gamma$,n)$^{196m,g}$Au, $^{183}$W($\gamma$,p)$^{182m,g}$Ta, and $^{195}$Pt($\gamma$,p)$^{194m,g}$Ir reactions, respectively, in the GDR region. For the calculation of the isomeric ratio, the authors applied a simple formula, using the statistical model proposed by Huizenga and Vandenbosch [17]. Though this model was established a long time ago, however, it is still a very powerful and widely used tool for nuclear reaction investigations. Comparison of calculations with the experiment showed that the isomeric ratios in the mentioned nuclei are extraordinarily low and they could be used to check nuclear reaction models. Such a discrepancy, unfortunately, is typical when measuring the *IR* values [18].

For a detailed analysis and testing different theoretical predictions, one needs to use the values of experimental cross-sections and isomeric ratios of multiparticle photonuclear reactions in a wide range of atomic mass and energies. However, there is still a lack of such experimental data [19 - 22].

The isomeric ratios *IR* for most photonuclear reactions change rapidly as the bremsstrahlung end-point energies increase from the reaction threshold up to the end of the GDR region. After 30 MeV the *IR* increases slowly or saturates in all energy ranges, as shown in the reviews [23, 24]. The fast-increasing isomeric ratios can be explained by a compound nuclear reaction mechanism in which the increased momentum was transferred to the compound nuclei. At higher incident energies the direct channel of the ($\gamma$,xn) or ($\gamma$,ypxn) reactions also occurs. The directly emitted particles carry away a relative-ly large angular momentum, and only a fraction of the energy and angular momentum of the incident quanta are transformed into the target nucleus. Above the energy of 30 MeV, the direct reaction channel dominates. The direct reactions largely suppress the population of states with higher spins, and the yield ratio of high to low spin states might not continue their rapidly increasing trend. These patterns are discussed in Refs. [24, 25].

This article presents the results of the *IR* measurements obtained at linear accelerator LUE-40 RDC "Accelerator" NSC KIPT for the target nuclei: Zr [26], Nb [27], Mo [28], Rh [29], Ag [30-32], In [30, 31], Sb [33-35], and Ta [36]. The calculations of theoretical *IR*s were performed using cross-sections of the TALYS1.95 code for six different level density models. In this article, when systematizing the results, we chose the representation of the isomeric ratio as the high spin to low spin states presentation.

## 1. EXPERIMENTAL SETUP AND PROCEDURE

We used two different schemes for performing studies of photonuclear reactions: in one of which an aluminum absorber is used to clean the bremsstrahlung flux from electrons, and in the second one, a deflecting magnet is used. The experimental Setups are schematically shown in Fig. 1 [37, 38].

### 1.1. EXPERIMENTAL SETUP 1

In the first experimental scheme, a converter, passing through which electrons interact with atoms of the substance and emit bremsstrahlung radiation, is used to obtain a flux of γ-quanta. To clean the bremsstrahlung γ-quanta flux from electrons, passed through the converter, a massive electron absorber consisting of light material (usually Al) [39, 40] is installed. This makes it possible to obtain an almost "pure" γ-ray flux on the target.

The advantages of this method are mainly in the simplicity of the approach. This guarantees a slight radiation and heat load of the target and, for example, on the components of the pneumatic tube transport, which is used for target delivery.

The disadvantages of this scheme are the distortion of the shape of the bremsstrahlung spectrum, and the additional generation of photoneutrons, which also contribute to studied reaction yield. In addition, in such a scheme of the experiment, modeling the bremsstrahlung flux in the GEANT4 [41] code is complicated.

### 1.2. EXPERIMENTAL SETUP 2

The second scheme of the experiment implements a deflecting magnet to divert the electrons that have passed through the converter [36] that allowing obtaining a "pure" beam of bremsstrahlung γ-quanta at the target [37]. This assumes the use of a thin Ta-converter to minimize the spread of the electron beam. When using the thin converter, the shape of the radiation spectrum can be described by a known analytical formula. At the same time, the influence of neutrons is minimized.

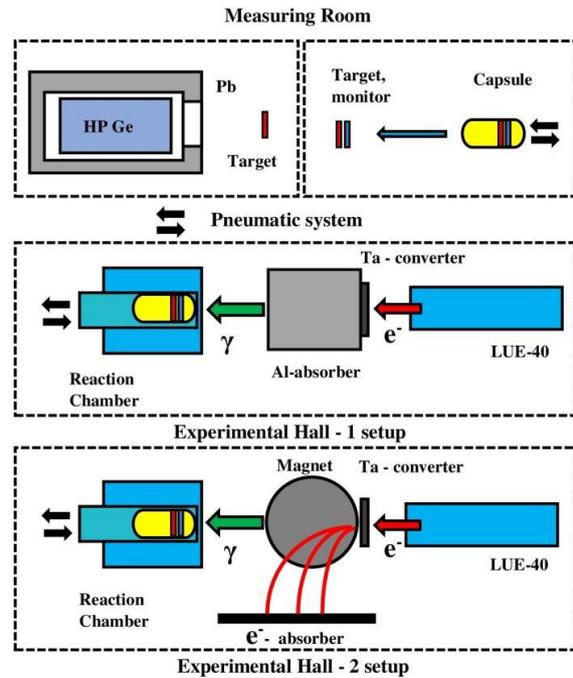

*Fig. 1. Experimental schematic diagram. Above – the measuring room, below – the experimental hall of the LUE-40 accelerator: Setup 1 – with Al-absorber; Setup 2 – with a deflecting magnet. Between the measuring room and the experimental hall, there is a pneumatic mail system*

### 1.3. EXPERIMENTAL PROCEDURE

The experiments were performed on the bremsstrahlung gamma-beam from the electron linear accelerator LUE-40 RDC "Accelerator" NSC KIPT using the method of induced γ-activity of the final product nucleus of the reaction. The experimental procedure is described in detail in [42 - 46].

The studies on the linac LUE-40 [47, 48] were performed at bremsstrahlung end-point energy range 30…100 MeV (this coincides with the energy of the electron beam $E_e$). The average current of the beam is $I_e \approx 3…4$ μA. The electron energy spectrum width at the full width at half maximum (FWHM) makes $\Delta E_e/E_e \sim 1\%$.

In the case of experimental Setup 1 – the bremsstrahlung gamma radiation was generated by passing a pulsed electron beam through a tantalum metal plate, 1.05 mm in thickness. The Ta-converter was fixed on an aluminum cylinder, 100 mm in diameter and 150 mm in thickness. For Setup 2 – the bremsstrahlung radiation was generated by passing an electron beam through the Ta foil, 0.1 mm in thickness and without Al-absorber.

In experiments for both Setup 1 and 2, the different targets, which represented thin discs with diameters 8…12 mm, were used.

For transporting the capsule with the sample to the place of irradiation and back for induced activity registration, a pneumatic tube transfer system was used. On delivery of the irradiated targets to the measuring room, the samples were extracted from the aluminum capsule and were transferred one by one to the detector for the measurements. Taking into account the time of target delivery and extraction from the capsule, the cooling time for the sample under study took no more than 3 minutes.

The γ-quanta of the reaction products were detected using a Canberra GC-2018 semiconductor HPGe detector with a relative detection efficiency of 20%. The resolution FWHM is 1.8 keV for energy $E_\gamma$ = 1332 keV and is 0.8 keV for $E_\gamma$ = 122 keV. The dead time for γ-quanta detection varied between 0.1…5%. The absolute detection efficiency for γ-quanta of different energies was obtained using a standard set of γ-quanta sources: $^{22}$Na, $^{60}$Co, $^{133}$Ba, $^{137}$Cs, $^{152}$Eu, $^{241}$Am.

The analytical curve in the form $\ln\varepsilon(E_\gamma) = \Sigma a_i(\ln E_\gamma)$, proposed in Ref. [49], was used to determine the value of $\varepsilon(E_\gamma)$ for various energies of γ-quanta.

## 2. CALCULATION FORMULAS

The cross-sections σ(E) of studied reactions for monochromatic photons were calculated using the TALYS1.95 open code [15] for the six different level density models *LD* 1-6. There are three phenomenological level density models and three options for microscopic level densities:

*LD*1: Constant temperature + Fermi gas model, introduced by Gilbert and Cameron [50]. In this model, the excitation energy range is divided into a low-energy part from $E_0$ up to a matching energy $E_M$, where the so-called constant temperature law applies, and a high-energy part above, where the Fermi gas model applies.

*LD*2: Back-shifted Fermi gas model [51], where the pairing energy is treated as an adjustable parameter and the Fermi gas expression is used down to $E_0$.

*LD*3: Generalized superfluid model (GSM) [52, 53]. The model takes superconductive pairing correlations into account according to the Bardeen-Cooper-Schrieffer theory.

*LD*4: Microscopic level densities (Skyrme force) from Goriely's tables [54]. Using this model allows reading tables of microscopic level densities from RIPL database [21]. These tables were computed by S. Goriely on based on Hartree-Fock calculations for excitation energies up to 150 MeV and for spin values up to $I$ = 30.

*LD*5: Microscopic level densities (Skyrme force) from Hilaire's combinatorial tables [55]. The combinatorial model includes a detailed microscopic calculation of the intrinsic state density and collective enhancement. The only phenomenological aspect of the model is a simple damping function for the transition from spherical to deformed.

*LD*6: Microscopic level densities based on temperature dependent Hartree-Fock-Bogoliubov calculations using the Gogny force [56] from Hilaire's combinatorial tables.

Using the theoretical cross-sections σ(E), one can obtain the reaction yield, which is determined by the formula:

$$Y(E_{\gamma\max}) = N_n \int_{E_{th}}^{E_{\gamma\max}} \sigma(E) W(E, E_{\gamma\max}) dE, \quad (1)$$

where $N_n$ is the number of atoms of the element under study; $W(E, E_{\gamma\max})$ is the bremsstrahlung γ-flux; $E_{thr}$ – the energy of the reaction threshold; $E_{\gamma\max}$ – the bremsstrahlung end-point energy.

The experimental values of the reaction yield were calculated as:

$Y(E_{\gamma\max}) = \lambda \Delta A /$
$(N_n I_\gamma \,\varepsilon\, (1 - \exp(-\lambda t_{irr})) \exp(-\lambda t_{cool})(1 - \exp(-\lambda t_{meas}))), \quad (2)$

where $I_\gamma$ – the intensity of the analyzed γ-quanta; ε – the absolute detection efficiency for the analyzed γ-quanta energy; λ is the decay constant ($\ln 2/T_{1/2}$); $\Delta A$ – the number of counts of γ-quanta in the full absorption peak; $t_{irr}$, $t_{cool}$ and $t_{meas}$ are the irradiation time, cooling time and measurement time, respectively.

In this work, the values of the isomeric ratio are calculated as the ratio of the yield $Y_H(E_{\gamma\max})$ of the formation of a nucleus in the High-level state to the yield $Y_L(E_{\gamma\max})$ of the formation of a nucleus in the Low-level state:

$$IR = Y_H(E_{\gamma\max})/Y_L(E_{\gamma\max}). \quad (3)$$

In the case of an internal transition between the metastable and ground states, it is necessary to take into account the accumulation of the ground state of the nucleus due to the decay of the metastable state. The formula for *IR* becomes more complex, for example, see [28, 57].

## 3. RESULTS AND DISCUSSIONS

The experimental isomeric ratios *IR* in (γ,*x*n) and (γ,p*x*n) reactions at the bremsstrahlung end-point energy range 30…100 MeV are systematized and presented below. These data were obtained at the LUE-40 linac RDC "Accelerator" NSC KIPT. Also, data from the international databases were used to compare experimental results and theoretical predictions.

The results obtained are presented in order of increasing atomic charge (mass): Zr [26], Nb [27], Mo [28], Rh [29], Ag [30 - 32], In [30, 31], Sb [33 - 35], Ta [36].

The article [26] presented the results from measuring and calculating the isomeric ratios for the $^{87m,g}$Y nucleus produced in the $^{90}$Zr(γ,2np)$^{87g,m}$Y and $^{91}$Zr(γ,3np)$^{87g,m}$Y reactions, and for the $^{86m,g}$Y nucleus produced in the $^{90}$Zr(γ,3np)$^{86g,m}$Y and $^{91}$Zr(γ,4np)$^{86g,m}$Y reactions at $E_{\gamma\max}$ = 84 MeV. In the work, isotopically enriched targets were used. The measured and calculated results of the isomeric ratio values *IR* are shown in Fig. 2.

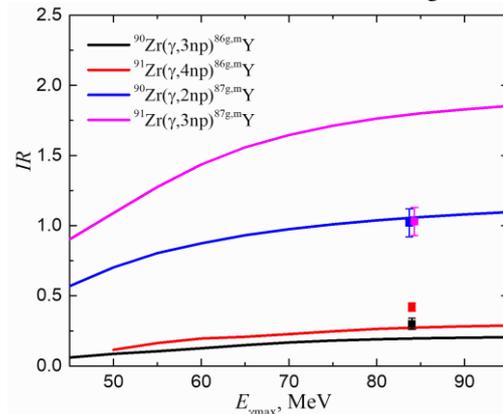

*Fig. 2. The measured and calculated results of the isomeric ratios IR for the $^{86m,g}$Y and $^{87m,g}$Y nucleus. Calculated in TALYS1.95, LD1 and experimental data [26] for different reactions are in the same colour*

In Ref. [27] the measured data of the isomeric ratios *IR* for the $^{93}$Nb(γ,4n)$^{89m,g}$Nb and $^{93}$Nb(γ,5n)$^{88m,g}$Nb reactions were presented. In this work we performed the *IR* calculations in TALYS1.95 code, *LD* 1-6. Experimental and theoretical results are shown in Figs. 3, 4.

The obtained values of *IR* for the $^{93}$Nb($\gamma$,4n)$^{89m,g}$Nb reaction are in satisfactory agreement with the results of other authors [58, 59]. The comparison shows that all theoretical estimates are higher than the experimental isomeric ratio for the $^{93}$Nb($\gamma$,4n)$^{89m,g}$Nb reaction (Fig. 3), and in good agreement for the case of the $^{93}$Nb($\gamma$,5n)$^{88m,g}$Nb reaction (Fig. 4).

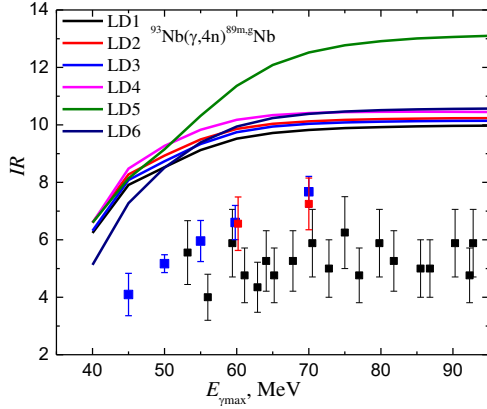

*Fig. 3. The measured and calculated results of the isomeric ratio IR for $^{93}$Nb($\gamma$,4n)$^{89m,g}$Nb reaction. Calculation in TALYS1.95, LD1-6. Experiment: black – [27], blue – [58], red – [59]*

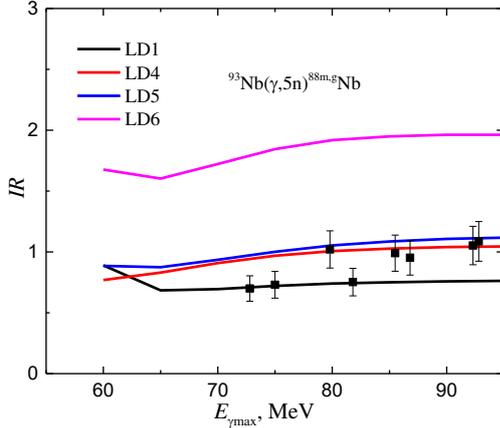

*Fig. 4. The measured and calculated results of the isomeric ratio IR for $^{93}$Nb($\gamma$,5n)$^{88m,g}$Nb reactions. Calculation in TALYS1.95, LD1, LD4-6. Experiment: black – [27]*

In work [28] the experimental isomeric ratio *IR* was determined for the reaction products $^{95m,g}$Nb at the bremsstrahlung end-point energy $E_{\gamma max}$ range of 38…93 MeV. The obtained values of *IR* are in satisfactory agreement with the results of other authors and extend the range of previously known data. The theoretical values of the yields $Y_{m,g}(E_{\gamma max})$ and the isomeric ratio *IR* for the isomeric pair $^{95m,g}$Nb from the $^{nat}$Mo($\gamma$,pxn) reaction were calculated using the partial cross-sections $\sigma(E)$ from the TALYS1.95 code for six different level density models. At the investigated range of $E_{\gamma max}$ the theoretical dependence of *IR* smoothly increases with increasing energy (Fig. 5).

The comparison showed a noticeable difference (more than 3.85 times) of the experimental isomeric ratio relative to all theoretical estimates in TALYS1.95 code.

The measured [29] and calculated in code TALYS1.95 results of the isomeric ratios *IR* for the $^{103}$Rh($\gamma$,4n)$^{99m,g}$Rh reaction are presented in Fig. 6.

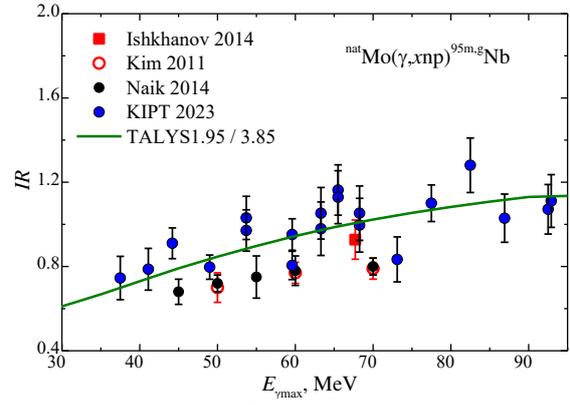

*Fig. 5. Isomeric ratio IR for the reaction products from the $^{nat}$Mo($\gamma$,xnp)$^{95m,g}$Nb reaction. Experimental results: blue circles – [28], square – [60], red empty circles – [59], black circles – [24]. The curve – calculation in the code TALYS1.95, LD3, divided by a factor of 3.85*

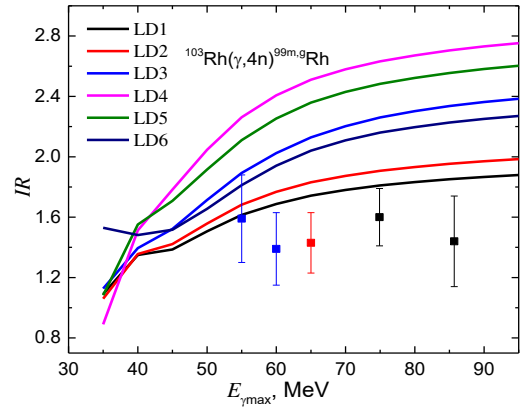

*Fig. 6. The measured and calculated results of the isomeric ratios IR for the $^{103}$Rh($\gamma$,4n)$^{99m,g}$Rh reaction. Calculation in TALYS1.95, LD1-6. Experiment: blue – [61], red – [62], black – [29]*

Values of the isomeric ratios for nuclei $^{104m,g}$Ag as products of the photonuclear reactions $^{107}$Ag($\gamma$,3n)$^{104m,g}$Ag and $^{109}$Ag($\gamma$,5n)$^{104m,g}$Ag were obtained in [30 - 32]. Natural silver consists of two stable isotopes: $^{107}$Ag – 0.51839 and $^{109}$Ag – 0.48161. Therefore, for the reaction $^{109}$Ag($\gamma$,5n)$^{104m,g}$Ag highly enriched targets $^{109}$Ag (with enrichment of more than 98%) were used.

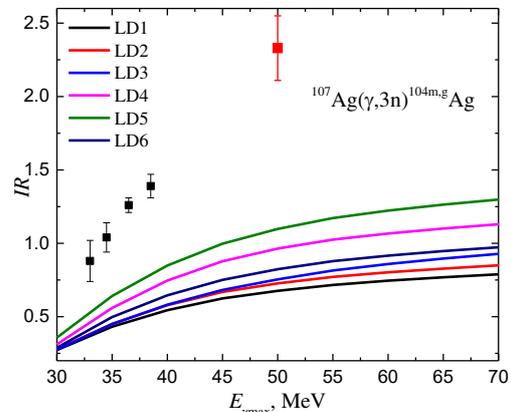

*Fig. 7. The measured and calculated results of the isomeric ratios IR for $^{107}$Ag($\gamma$,3n)$^{104m,g}$Ag reaction. Calculation in TALYS1.95, LD 1-6. Experiment: black – [30 - 32], red – [63]*

The measured and calculated results of the isomeric ratios *IR* are shown in Fig. 7 for $^{107}$Ag(γ,3n)$^{104m,g}$Ag reaction and in Fig. 8 for $^{109}$Ag(γ,5n)$^{104m,g}$Ag reaction.

The isomeric ratios for nuclei $^{110m,g}$In and $^{108m,g}$In as products of the photonuclear reactions $^{113}$In(γ,3n)$^{110m,g}$In, $^{115}$In(γ,5n)$^{110m,g}$In and $^{115}$In(γ,7n)$^{108m,g}$In were obtained in [30 - 32]. Natural In consists of two stable isotopes: $^{113}$In – 0.0429 and $^{115}$In – 0.9571. For the reactions $^{115}$In(γ,5n)$^{110m,g}$In, $^{115}$In(γ,7n)$^{108m,g}$In highly enriched targets $^{115}$In (with enrichment more than 99 %) were used.

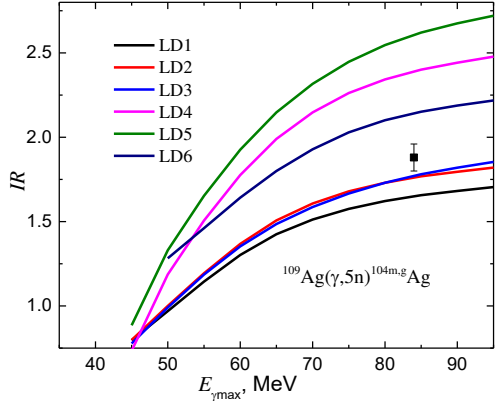

*Fig. 8. The measured and calculated results of the isomeric ratios IR for $^{109}$Ag(γ,5n)$^{104m,g}$Ag reaction. Calculation in TALYS1.95, LD 1-6. Experiment: black – [30 - 32]*

The measured and calculated results of the isomeric ratios *IR* shown in Fig. 8 for the $^{113}$In(γ,3n)$^{110m,g}$In reaction, $^{115}$In(γ,5n)$^{110m,g}$In reaction, and in Fig. 9 for the $^{115}$In(γ,7n)$^{108m,g}$In reaction.

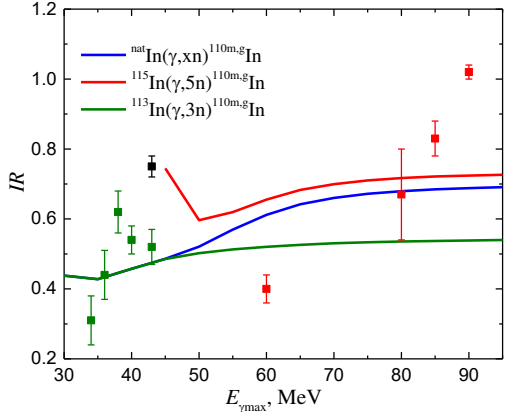

*Fig. 9. The measured and calculated results of the isomeric ratios IR for $^{113}$In(γ,3n)$^{110m,g}$In (green) and $^{115}$In(γ,5n)$^{110m,g}$In (red) reactions. Calculation in TALYS1.95, LD1. Experiment: green – $^{113}$In(γ,3n)$^{110m,g}$In [30, 31], red – $^{115}$In(γ,5n)$^{110m,g}$In [30 - 32], black – [64]*

The $^{113}$In(γ,3n)$^{110m,g}$In reaction is separated from the $^{115}$In(γ,5n)$^{110m,g}$In reaction by the reaction threshold. Fig. 9 shows calculations of the *IR* for reaction $^{115}$In(γ,5n)$^{110m,g}$In for the $^{115}$In and $^{nat}$In targets. In the case of a $^{nat}$In target, there is a contribution from the $^{113}$In(γ,3n)$^{110m,g}$In reaction which, according to calculations in the TALYS code, decreases smoothly from 78 to 17% in the energy range 50...95 MeV. That is why enriched targets were used.

Note that calculations give different *IR* values for reactions with different indium isotopes. As in the case of the formation of the $^{95m,g}$Nb isomer pair in the $^{nat}$Mo(γ,pxn) reaction on four stable Mo isotopes [28].

Fig. 10 shows calculations of the *IR* of the $^{115}$In(γ,7n)$^{108m,g}$In reaction for the $^{115}$In and $^{nat}$In targets. In the case of a $^{nat}$In target, there is a contribution from the $^{113}$In(γ,5n)$^{108m,g}$In reaction, which smoothly drops from 98 to 19% in the energy range 50...95 MeV.

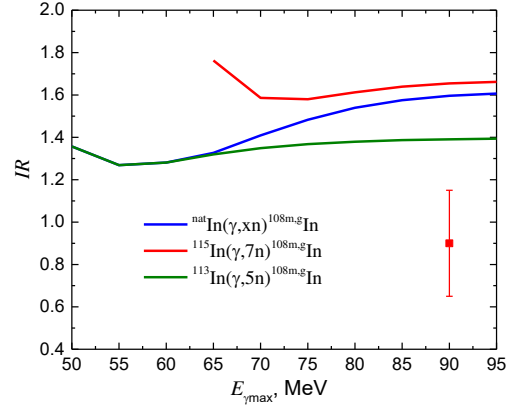

*Fig. 10. The measured and calculated results of the isomeric ratios IR for $^{113}$In(γ,5n)$^{108m,g}$In (green) and $^{115}$In(γ,7n)$^{108m,g}$In (red) reactions. Calculation in TALYS1.95, LD1. Experiment: red – $^{115}$In(γ,7n)$^{108m,g}$In [30 - 32]*

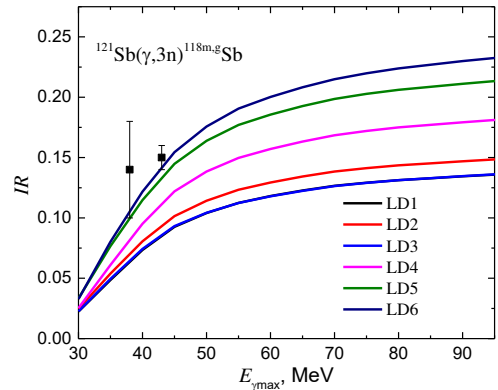

*Fig. 11. The measured and calculated results of the isomeric ratio IR for $^{121}$Sb(γ,3n)$^{118m,g}$Sb reaction. Calculation in TALYS1.95, LD1-6. Experiment: black – [33 - 35]*

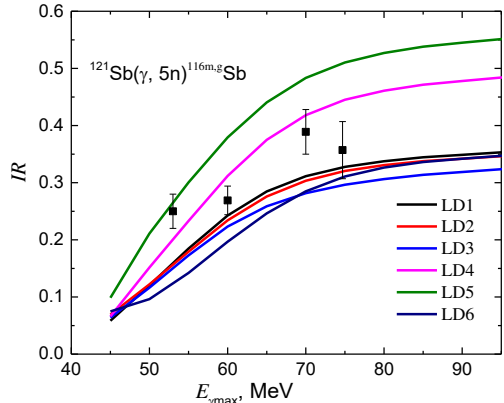

*Fig. 12. The measured and calculated results of the isomeric ratios IR for $^{121}$Sb(γ,5n)$^{116m,g}$Sb reaction. Calculation in TALYS1.95, LD1-6. Experiment: black – [33 - 35]*

In [33 - 35] the experimental values of the isomeric ratios were obtained for $^{118m,g}$Sb and $^{116m,g}$Sb nuclei as products of the photonuclear reactions $^{121}$Sb($\gamma$,3n)$^{118m,g}$Sb and $^{121}$Sb($\gamma$,5n)$^{116m,g}$Sb, using bremsstrahlung end-point energies in the region from 33 to 53 MeV (Figs. 11 and 12). To correctly derive the isomeric ratio for $^{116m,g}$Sb we took into account the contribution of $^{116m,g}$In from reaction $^{121}$Sb($\gamma$,n$\alpha$)$^{116m,g}$In.

Experimental values of isomeric ratios for $^{118m,g}$Sb from the reaction $^{121}$Sb($\gamma$,3n)$^{118m,g}$Sb and isomeric ratios for $^{116m,g}$Sb from the reaction $^{121}$Sb($\gamma$,5n)$^{116m,g}$Sb are in a sufficiently good agreement with theoretical calculation results.

Fig. 13 shows the experimental and theoretical values of the isomeric ratio *IR* of the reaction products $^{181}$Ta($\gamma$,3n)$^{178m,g}$Ta [65].

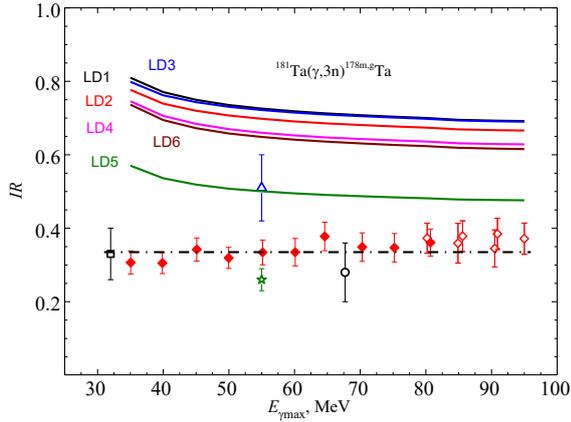

*Fig. 13. The measured and calculated results of the isomeric ratio IR of the $^{181}$Ta($\gamma$,3n)$^{178m,g}$Ta reaction. Experiment: full rhombuses – [36], empty rhombuses – [65], star – [66], circle – [67], square – [68], triangle – [69]. Solid lines – calculations of IR using TALYS1.95, LD1-6, dash-dotted line – approximation of all experimental data by a horizontal line (0.335±0.008)*

The experimental results obtained on the LUE-40 linac RDC "Accelerator" NSC KIPT are presented in Table.

*Isomeric ratio IR = $Y_H(E_{\gamma max})/Y_L(E_{\gamma max})$ obtained in the experiments on LUE-40 linac RDC "Accelerator" NSC KIPT*

| Reaction, Jm, Jg | $E_{\gamma max}$, MeV | IR ± ΔIR |
|---|---|---|
| $^{90}$Zr($\gamma$,2np)$^{87g,m}$Y, m: 9/2+ g: 1/2- | 84.0 | 1.02 ±0.10 |
| $^{91}$Zr($\gamma$,3np)$^{87g,m}$Y, m: 9/2+ g: 1/2- | 84.0 | 1.03 ±0.10 |
| $^{90}$Zr($\gamma$,2np)$^{86g,m}$Y, m: 8+ g: 4- | 84.0 | 0.30 ±0.04 |
| $^{91}$Zr($\gamma$,3np)$^{86g,m}$Y, m: 8+ g: 4- | 84.0 | 0.42 ±0.03 |
| $^{93}$Nb($\gamma$,4n)$^{89m,g}$Nb, m: 1/2- g: 9/2+ | 53.2 | 5.56±1.11 |
| | 56.0 | 4.00±0.80 |
| | 59.4 | 5.88±1.18 |
| | 61.1 | 4.76±0.95 |
| | 62.9 | 4.35±0.87 |
| | 64.1 | 5.26±1.05 |
| | 65.2 | 4.76±0.95 |
| | 67.8 | 5.26±1.05 |
| | 70.5 | 5.88±1.18 |
| | 72.8 | 5.00±1.00 |
| | 75.0 | 6.25±1.25 |
| | 77.0 | 4.76±0.95 |
| | 79.8 | 5.88±1.18 |
| | 81.8 | 5.26±1.05 |
| | 85.5 | 5.00±1.05 |
| | 86.8 | 5.00±1.00 |
| | 90.3 | 5.88±1.18 |
| | 92.3 | 4.76±0.95 |
| | 92.8 | 5.88±1.18 |
| $^{93}$Nb($\gamma$,5n)$^{88m,g}$Nb, m: 4- g: 8+ | 72.8 | 0.70±0.10 |
| | 75.0 | 0.73±0.11 |
| | 79.8 | 1.02±0.15 |
| | 81.8 | 0.75±0.11 |
| | 85.5 | 0.99±0.15 |
| | 86.8 | 0.95±0.14 |
| | 92.3 | 1.05±0.16 |
| | 92.8 | 1.09±0.16 |
| $^{nat}$Mo($\gamma$,xnp)$^{95m,g}$Nb m: 1/2- g: 9/2+ | 37.50 | 0.75 ± 0.10 |
| | 41.10 | 0.79 ± 0.10 |
| | 44.20 | 0.91 ± 0.07 |
| | 49.00 | 0.80 ± 0.06 |
| | 53.70 | 0.97 ± 0.10 |
| | 53.70* | 1.03 ± 0.10 |
| | 59.60 | 0.95 ± 0.07 |
| | 59.60* | 0.81 ± 0.07 |
| | 63.30 | 1.05 ± 0.12 |
| | 63.30* | 0.98 ± 0.13 |
| | 65.50 | 1.16 ± 0.12 |
| | 65.50* | 1.13 ± 0.13 |
| | 68.25 | 1.05 ± 0.13 |
| | 68.25* | 1.00 ± 0.13 |
| | 73.10 | 0.83 ± 0.11 |
| | 77.50 | 1.10 ± 0.09 |
| | 82.50 | 1.28 ± 0.13 |
| | 86.90 | 1.03 ± 0.11 |
| | 92.50 | 1.07 ± 0.12 |
| | 92.90 | 1.11 ± 0.13 |
| $^{103}$Rh($\gamma$,4n)$^{99m,g}$Rh m: 9/2+ g: 1/2- | 74.9 | 1.60±0.19 |
| | 85.7 | 1.44±0.30 |
| $^{107}$Ag($\gamma$,3n)$^{104m,g}$Ag, m: 2+ g: 5+ | 33.0 | 0.88 ±0.14 |
| | 34.5 | 1.04 ±0.10 |
| | 36.5 | 1.26±0. 05 |
| | 38.5 | 1.39±0. 08 |
| $^{109}$Ag($\gamma$,5n)$^{104m,g}$Ag, m: 2+ g: 5+ | 84.0 | 1.88 ±0.08 |
| $^{113}$In($\gamma$,3n)$^{110m,g}$In, m: 2+ g: 7+ | 32.0 | 0.31 ±0.07 |
| | 34.0 | 0.44 ±0.07 |
| | 36.0 | 0.62 ±0.06 |
| | 38.5 | 0.54 ±0.04 |
| | 41.5 | 0.52 ±0.05 |
| $^{115}$In($\gamma$,5n)$^{110m,g}$In, m: 2+ g: 7+ | 56.0 | 0.4 ± 0.04 |
| | 75.0 | 0.67 ±0.13 |
| | 79.0 | 0.83 ±0.08 |
| | 84.0 | 1.02 ±0.02 |
| $^{115}$In($\gamma$,7n)$^{108m,g}$In, m: 2+ g: 7+ | 84.0 | 0.90 ± 0.25 |
| $^{121}$Sb($\gamma$,3n)$^{118m,g}$Sb, m: 8- g: 1+ | 38.0 | 0.14 ±0.04 |
| | 43.0 | 0.15± 0.01 |
| $^{121}$Sb($\gamma$,5n)$^{116m,g}$Sb, m: 8- g: 3+ | 53.0 | 0.250±0.030 |
| | 60.0 | 0.269±0.025 |
| | 70.0 | 0.389±0.039 |
| | 74.7 | 0.357±0.050 |

| $^{181}$Ta(γ,3n)$^{178m,g}$Ta m: 1+ g: 7- | 35.1 | 0.307±0.032 |
|---|---|---|
| | 39.9 | 0.306±0.029 |
| | 45.1 | 0.340±0.032 |
| | 50.0 | 0.320±0.029 |
| | 55.2 | 0.330±0.034 |
| | 60.1 | 0.335±0.037 |
| | 64.6 | 0.378±0.039 |
| | 70.3 | 0.349±0.039 |
| | 75.2 | 0.347±0.039 |
| | 80.7 | 0.361±0.037 |
| | 80.2 | 0.373±0.041 |
| | 85.6 | 0.379±0.042 |
| | 90.9 | 0.385±0.042 |
| | 95.0 | 0.372±0.043 |
| | 84.9 | 0.359±0.054 |
| | 90.5 | 0.345±0.050 |

## CONCLUSIONS

The systematization of the results, obtained at the electron beam of the LUE-40 linac RDC "Accelerator" NSC KIPT, and calculations in the TALYS1.95 code showed that experimental data for *IR* do not always agree with calculations, especially in the case of multi-particle reactions with the escape of charged particles.

For a more detailed comparison of experimental data and theoretical estimates, it is preferable to perform direct measurements of the cross-sections for nuclear formation in the ground and metastable states. This allows one to analyze the energy dependencies of reaction cross-sections and yields. Unfortunately, in some cases, for example, in the presence of an internal transition, direct measurement of the cross-section is not available. Therefore, the *IR* measurements and their analysis are a good alternative.

As the energy $E_{\gamma max}$ increases, the cross-sections of photonuclear reactions are integrated and, accordingly, the *IR* value reaches saturation. However, a simple analytical formula for estimating *IR* currently does not exist. In addition, for high-threshold reactions, in some cases, saturation of the *IR* does not occur in the energy range under study.

Two important fields of the *IR* research can be distinguished. First, the analysis of the *IR* values, obtained for one final isomeric pair, but in different reactions, i.e. from different initial nuclei. There are different *IR* values for the same final state obtained by different types of reactions (γ,*x*), (p,*x*), (n,*x*), (α,*x*), etc. Second, analysis of the differences in *IR* obtained from different isotopes of the element. Experiments and calculations in TALYS show that differences in *IR* occur in the case of photonuclear reactions for different isotopes of the same element during the transition to one final nucleus. This behavior of *IR* for isotopes is associated with different values of the initial spin of the nucleus under study and different nucleon compositions of the compound nucleus. This makes it interesting to carry out measurements on monoisotopes such as indium or molybdenum.

Further directions of experimental research at NSC KIPT may include:

1. multiparticle high-threshold photoneutron reactions (γ,*x*n);

2. photonuclear reactions with protons and charged particles in the output channel (γ,*y*p*x*n);

3. the experiments on different isotopes (Mo, Sb, Ag, etc.) of a chemical element in order to determine differences in the *IR* for the same final isomeric pair.

Despite the fact that the cross sections for multiparticle reactions and reactions with charged particles in the output channel, as a rule, have small values, the existing experimental facilities (accelerator, two experimental setups, pneumatic mail system, etc.) allow such studies.

## REFERENCES

1. S.Y.F. Chu, L.P. Ekstrom, R.B. Firestone. The Lund/LBNL, Nuclear Data Search, Version 2.0, February 1999. WWW Table of Radioactive Isotopes, available from http://nucleardata.nuclear.lu.se/toi/.
2. R. Volpel. Bestimmung von Isomerenverhältnissen bei Verschiedenen Kernphoto- und (n,2n)-Reaktionen // *Nucl. Phys.* 1972, A182, p. 411, doi.org/10.1016/0375- 9474(72)90287-4.
3. H. Bartsch, K. Huber, U. Kneissl, and H. Krieger, Critical consideration of the statistical model analysis of photonuclear isomeric cross-section ratios // *Nucl. Phys.* 1976, A256, p. 243. doi.org/10.1016/0375-9474(76)90106-8.
4. D. Kolev, E. Dobreva, N. Nenov, and V. Todorov. A convenient method for experimental determination of yields and isomeric ratios in photonuclear reactions measured by the activation technique // *Nucl. Instrum. Method. Phys. Research.* 1995, A356, p. 390. doi.org/10.1016/0168-9002(94)01319-5.
5. I.B. Haller, G. Rudstam. Relative yields of the isomeric pairs $^{69g}$Zn, $^{69m}$Zn and $^{52g}$Mn, $^{52m}$Mn in some spallation reactions induced by 20-153 MeV protons // *J. Inorg. Nucl. Chem.* 1961, 19, p. 1, doi.org/10.1016/0022-1902(61)80038-9.
6. D. Kolev. Studies of some isomeric yield ratios produced with bremsstrahlung // *Appl. Radiat. Isot.* 1998, 49, p. 989. doi.org/10.1016/S0969-8043(97)10094-X.
7. J.R. Huizenga and R. Vandenbosch. Interpretation of Isomeric Cross-Section Ratios for (n,γ) and (γ,n) Reactions // *Phys. Rev.* 1960, 120, p. 1305, doi.org/10.1103/PhysRev.120.1305.
8. R. Vandenbosch and J.R. Huizenga. Isomeric Cross Section Ratios for Reactions Producing the Isomeric Pair Hg197, 197m // *Phys Rev* 1960, 12, p. 1313, doi.org/10.1103/PhysRev.120.1313.
9. H.A. Bethe. Nuclear Physics B. Nuclear Dynamics, Theoretical // *Rev. Mod. Phys.* 1937, 9, p. 84, doi.org/10.1103/RevModPhys.9.69.
10. C. Bloch. Theory of Nuclear Level Density // *Phys. Rev.* 1954, 93, p. 1094. doi.org/10.1103/PhysRev. 93.1094.
11. K.J. Le Couteur, D.W. Lang. Neutron evaporation and level densities in excited nuclei // *Nucl. Phys.* 1959, 13, p. 32. doi.org/10.1016/0029-5582(59)90136-1.
12. M.B. Chadwick, P. Oblozinsky, P.E. Hodgson, and G. Reffo. Pauli-blocking in the quasideuteron model of photoabsorption // *Phys. Rev.* 1991, C44, p. 814. doi.org/10.1103/PhysRevC.44.814.

13. B.S. Ishkhanov, V.N. Orlin, and S.Yu. Troschiev, Photodisintegration of Tantalum // *Phys. Atom. Nucl.* 2012, 75, p. 253, doi.org/10.1134/S1063778812020093. (Yadernaya Fizika. 2012, 75, p. 283).
14. A. Rodrigo, N. Otuka, S. Tak´acs, A.J. Koning. Compilation of isomeric ratios of light particle induced nuclear reactions // *Atom. Data Nucl. Data Tables.* 2023, 153, p. 101583, doi.org/10.1016/j.adt.2023.101583.
15. A. Koning and D. Rochman // *Nucl. Data Sheets* 2012, v. 113, p. 2841, TALYS – based evaluated nuclear data library. https://tendl.web.psi.ch/tendl2019/tendl2019.html.
16. T.D. Thiep, T.Th. An, Ph.V. Cuong, et al. The study of isomeric ratios in photonuclear reactions forming high spin isomers in the giant dipole resonance region // *Communications in Physics.* 2014, 24, p. 381-389. doi.org/10.15625/0868-3166/24/4/5553.
17. J.R. Huizenga, R. Vandenbosch. Interpretation of Isomeric Cross-Section Ratios for (n, γ) and (γ, n) Reactions // *Phys. Rev.* 1960, 120, p. 1306. doi.org/10.1103/PhysRev.120.1305.
18. A.G. Belov, Yu.P. Gangrskyi, N.N. Kolesnikov, et al. Excitation of high-spin isomers in photonuclear reactions // *Letters of EhChAYa.* 2004, 1, p. 47-52.
19. Experimental Nuclear Reaction Data (EXFOR) // https://www-nds.iaea.org/exfor/.
20. Data Center of Photonuclear Experiments, http://cdfe.sinp.msu.ru/.
21. R. Capote et al. RIPL – Reference Input Parameter Library for calculation of nuclear reactions and nuclear data evaluation // *Nucl. Data Sheets.* 2009, 110, 3107. https://wwwnds.iaea.org/RIPL/.
22. S.Y.F. Chu, L.P. Ekstrom, R.B. Firestone. The Lund/LBNL, Nuclear Data Search, Version 2.0, February 1999, WWW Table of Radioactive Isotopes. http://nucleardata.nuclear.lu.se/toi/.
23. O.M. Vodin, O.A. Bezshyyko, L.O. Golinka-Bezshyyko, et al. Isomer ratios in photonuclear reactions with multiple neutron emission // *Problems of Atomic Science and Technology.* 2019, № 3, p. 38-46.
24. H. Naik, G. Kim, Kw. Kim, M. Zaman, et al. Independent isomeric yield ratios of $^{95m,g}$Nb in the $^{nat}$Mo(γ,xnp) and $^{nat}$Zr(p, xn) reactions // *J. Radioanal. Nucl. Chem.* 2014, 300, p. 1121-1130. doi.org/10.1007/s10967-014-3045-x.
25. N.V. Do, P.D. Khue, K.T. Thanh, et al. Isomeric yield ratios for the $^{nat}$Sb(γ,xn)$^{120m,g,122m,g}$Sb reactions measured at 40-, 45-, 50-, 55-, and 60-MeV bremsstrahlung energies // *Nucl. Instr. Meth. Phys. Research.* 2012, B283, p. 40-45. dx.doi.org/10.1016/j.nimb.2012.04.015.
26. O.A. Bezshyyko, A.N. Vodin, L.A. Golinka-Bezshyyko, et al. Isomeric Ratios of the Products of (γ,xnp) Reactions on $^{90,91}$Zr Nuclei for a Maximum Bremsstrahlung Energy of 84 MeV // *Bulletin of the Russian Academy of Sciences. Physics.* 2011, 75, p. 937-940. doi.org/10.3103/ S1062873811070094.
27. A.N. Vodin, O.S. Deiev, V.Yu. Korda, et al. Photoneutron reactions on $^{93}$Nb at E$_{γmax}$ = 33-93 MeV // *Nucl. Phys.* 2021, A1014, p. 122248. doi.org/10.1016/j.nuclphysa.2021.122248, arXiv: 2101.08614.
28. I.S. Timchenko, O.S. Deiev, S.M. Olejnik, et al. Isomeric pair $^{95m,g}$Nb in the photonuclear reactions on $^{nat}$Mo at the bremsstrahlung end-point energy of 38-93 MeV // *Chin. Phys.* 2023, C47(12). doi.org/10.1088/1674-1137/acfaed, arXiv:2308.02243.
29. O. Bezshyyko, A.N. Vodin, L. Golinka-Bezshyyko, et al. Isomer ratios for products of photonuclear reactions on Rh // *EPJ Web of Conferences.* 2020, 239, p. 01026. doi.org/10.1051/epjconf/202023901026.
30. O.A. Bezshyyko, A.N. Vodin, L.A. Golinka-Bezshyyko, et al. Isomer ratios of products from photonuclear reactions on silver and indium nuclei at γ ray energies above 35 MeV // *Bulletin of the Russian Academy of Sciences: Physics.* 2009, 73, p. 1461. doi.org/10.3103/S1062873809110070.
31. O.A. Bezshyyko, A.N. Vodin, L.A. Golinka-Bezshyyko, et al. Isomer ratios for products of photonuclear reactions with middle-weight nuclei // *Bulletin of the Russian Academy of Sciences: Physics.* 2011, 75, p. 941-945. doi.org/10.3103/S1062873811070100.
32. O.A. Bezshyyko, A.N. Vodin, L.O. Golinka-Bezshyyko, A.N. Dovbnya, et al. Isomer ratios for product of photonuclear reaction 107Ag(γ;3n)104m;gAg in the energy region 35÷40 MeV // *Nuclear Physics and Atomic Energy.* 2009, v. 10, p. 61-65.
33. O. Bezshyyko, A. Dovbnya, L. Golinka-Bezshyyko, et al. Isomer ratios for products of photonuclear reactions on $^{121}$Sb // *EPJ Web of Conferences.* 2017, 146, p. 05016. doi.org/10.1051/epjconf/201714605016.
34. V.I. Berest, O.A. Bezshiyko, O.M. Vodin, et al. Isomeric ratios in the reaction $^{121}$Sb(γ,5n)$^{116m,g}$Sb // *IEP-2021*, Institute of Electronic Physics of the National Academy of Sciences of Ukraine, BOOK OF ABSTRACTS, International Conference of Young Scientists and Post-Graduate Students, Uzhhorod, 26-28 May 2021, p. 54-55.
35. O.A. Bezshyiko, O. Bezshyyko, L. Golinka-Bezshyyko, et al. Isomer Ratios for Products of Photonuclear Reactions $^{121}$Sb(γ,3n)$^{118m,g}$Sb and $^{121}$Sb(γ,5n)$^{116m,g}$Sb // *WDS'12 Proceedings of Contributed Papers, Part III.* 2012, p. 147-151.
36. O.S. Deiev, I.S. Timchenko, S.N. Olejnik, et al. Isomeric ratio of the $^{181}$Ta(γ,3n)$^{178m,g}$Ta reaction products at energy E$_{γmax}$ up to 95 MeV // *Chin. Phys. C.* 2022, v. 46, № 1, p. 014001. doi.org/10.1088/ 1674-1137/ac2a95.
37. V.V. Mytrochenko, L.I. Selivanov, V.Ph. Zhyglo, et al. Magnetic system for cleaning the gamma beam at the LUE-40 linac output // *Problems of Atomic Science and Technology. Series "Nuclear Physics Investigations".* 2022, v. 3, № 139, p. 62-67, doi.org/10.46813/2022-139-062.
38. O.S. Deiev, I.S. Timchenko, S.M. Olejniket, et al. Photonuclear reactions cross-section at energies up to 100 MeV for different experimental setups // *Problems of Atomic Science and Technology.* 2022, № 5, p. 11-18. doi.org/10.46813/2022-141-011.
39. S.S. Belyshev, A.N. Ermakov, B.S. Ishkhanov, et al. Studying photonuclear reactions using the activation technique // *Nucl. Instrum. Meth. Phys. Research.* 2014, A.745, p. 133-137.

40. L.M. Young. *Photoneutron Cross Sections and Spectra from Monoenergetic Photons on Yttrium, Praseodimium, Lead, and Bismuth in the Giant Resonance*: Ph.D. Thesis, University of Illinois, USA. 1972.
41. S. Agostinelli, J. Allison, K. Amako, et al. Geant4— a simulation toolkit // *Nucl. Instrum. Meth.* 2003, v. A506, p. 250, https://doi.org/10.1016/S0168-9002(03)01368-8.
42. O.S. Deiev, I.S. Timchenko, S.M. Olejnik, et al. Cross-sections of photoneutron reactions on $^{181}$Ta at $E_{\gamma max}$ up to 95 MeV // *Phys. Rev.* 2022, C106, p. 024617. doi.org/10.1103/ PhysRevC.106.024617.
43. A.N. Vodin, O.S. Deiev, I.S. Timchenko, S.N. Olejnik. Cross-sections for the $^{27}$Al($\gamma$,x)$^{24}$Na multiparticle reaction at $E_{\gamma max}$ = 35-95 MeV // *Eur. Phys. J.* 2021, A57, p. 207. arXiv:2012.14475, doi.org/10.1140/epja/s10050-021-00483-y.
44. O.S. Deiev, I.S. Timchenko, S.N. Olejnik, et al. Cross-sections for the $^{27}$Al($\gamma$,x)$^{22}$Na multichannel reaction with the 28.3 MeV difference of the reaction thresholds // *Chin. Phys.* 2022, C46, p. 064002. doi.org/10.1088/1674-1137/ac5733, arXiv:2105.12658.
45. O.S. Deiev, I.S. Timchenko, S.N. Olejnik, et al Photonuclear reactions $^{65}$Cu($\gamma$,n)$^{64}$Cu and $^{63}$Cu($\gamma$,xn)$^{63-x}$Cu. cross-sections in the energy range $E_{\gamma max}$ = 35-94 MeV // *Chin. Phys.* 2022, C460, p. 124001. doi.org/10.1088/1674-1137/ac878a.
46. A.N. Vodin, O.S. Deiev, I.S. Timchenko, S.O. Perezhogin, V.O. Bocharov. Cross-sections of photonuclear reactions on $^{nat}$Mo targets at end-point bremsstrahlung energy up to $E_{\gamma max}$ = 100 MeV // *Problems of Atomic Science and Technology.* 2021, № 3, p. 98-103.
47. A.N. Dovbnya, M.I. Aizatsky, V.N. Boriskin, et al. Beam parameters of an S-band electron linac with beam energy of 30…100 MeV // *Problems of Atomic Science and Technology.* 2006, № 2, p. 11.
48. M.I. Aizatskyi, V.I. Beloglazov, V.N. Boriskin, et al. State and prospects of the linac of nuclear-physics complex with energy of electrons up to 100 MeV // *Problems of Atomic Science and Technology.* 2014, № 3, p. 60.
49. G.L. Molnar, Zs. Revay, T. Belgya. Wide energy range efficiency calibration method for Ge detectors // *Nucl. Instrum. Methods Phys. Res.* 2002, A489, p. 140. doi.org/10.1016/S0168-9002(02)00902-6.
50. A. Gilbert and A.G.W. Cameron. A composite nuclear level density formula with shell corrections // *Canad. Journ. Phys.* 1965, 43, p. 1446, doi.org/10.1139/p65-139.
51. W. Dilg, W. Schantl, H. Vonach, and M. Uhl. Level density parameters for the back-shifted fermi gas model in the mass range 40 < A < 250 // *Nucl. Phys.* 1973, A217, p. 269; doi.org/10.1016/0375-9474(73)90196-6.
52. A.V. Ignatyuk, K.K. Istekov, and G.N. Smirenkin. Role of collective effects in the systematics of nuclear level densities // *Sov. J. Nucl. Phys.* 1979, 29, p. 450.
53. A.V. Ignatyuk, J.L. Weil, S. Raman, and S. Kahane. Density of discrete levels in $^{116}$Sn // *Phys. Rev.* 1993, C47, p. 1504. doi.org/10.1103/PhysRevC.47.1504.
54. S. Goriely, F. Tondour, and J.M. Pearson. A Hartree-Fock nuclear mass table // *Atom. Data Nucl. Data Tables.* 2001, 77, p. 311. doi.org/10.1006/adnd.2000.0857.
55. S. Goriely, S. Hilaire, and A.J. Koning. Improved microscopic nuclear level densities within the Hartree-Fock-Bogoliubov plus combinatorial method // *Phys. Rev.* 2008, C78, p. 064307. doi.org/10.1103/PhysRevC.78.064307.
56. S. Hilaire, M. Girod, S. Goriely, and A.J. Koning. Temperature-dependent combinatorial level densities with the D1M Gogny force // *Phys. Rev.* 2012, C86, p. 064317. doi.org/10.1103/PhysRevC.86.064317.
57. S.R. Palvanov, O. Razhabov, M. Kajumov, et al. Isomeric yield ratios of the ($\gamma$,n) and ($\gamma$,2n) reactions on nuclei of $^{110}$Pb, $^{142}$Nd, and $^{144}$Sm // *Bulletin Rus. Acad. Sciences. Physics.* 2011, 75, p. 222. (Izvestiya Ross. Akad. Nauk. 2011, 75, p. 239.
58. H. Naik, K.S. Kim, G. Kim, et al. Independent isomeric-yield ratio of $^{89m,g}$Nb from $^{93}$Nb($\gamma$;4n), $^{nat}$Zr(p;xn), and $^{89}$Y($\gamma$;4n) reactions // *Journal of Radioanalytical and Nuclear Chemistry.* 2014, v. 299, p. 1335-1343.
59. K.S. Kim, MD. Sh. Rahman, M. Lee, G. Kim, et al. Measurement of isomeric yield ratios for $^{93}$Nb($\gamma$,4n)$^{89m,g}$Nb and $^{nat}$Mo($\gamma$,xnp)$^{95m,g}$Nb reactions with 50-, 60-, and 70-MeV bremsstrahlung // *J. Radioanal. Nucl. Chem.* 2011, v. 287, p. 869-877, doi.org/10.1007/s10967-010-0839-3.
60. B.S. Ishkhanov, I.M. Kapitonov, A.A. Kuznetsov, et al. Photonuclear Reactions on Molybdenum Isotopes // *Physics of Atomic Nuclei.* 2014, v. 77, № 11, p. 1362-1370. (Yadernaya Fizika, 2014, v. 77, № 11, p. 1427-1435.) doi.org/10.1134/S106377881410007X.
61. M.Sh. Rahman, K. Kim, G. Kim, et al. Measurement of flux-weighted average cross-sections and isomeric yield ratios for $^{103}$Rh($\gamma$,xn) reactions in the bremsstrahlung end-point energies of 55 and 60 MeV // *Eur. Phys. J.* 2016, A52, p. 194. doi.org/10.1140/epja/i2016-16194-x.
62. Van Do Nguyen, Duc Khue Pham, Tien Thanh Kim, et al. Measurement of Isomeric Cross-Section Ratios for the $^{45}$Sc($\gamma$,n)$^{44m,g}$Sc, $^{nat}$Ti($\gamma$,x)$^{44m,g}$Sc, $^{103}$Rh($\gamma$,4n)$^{99m,g}$Rh, and $^{nat}$Fe($\gamma$,x)$^{52m,g}$Mn Reactions Induced by 65-MeV Bremsstrahlung // *Journal of the Korean Physical Society.* 2007, 50, p. 417.
63. A. Ermakov, B. Ishkhanov, I. Kapitonov, et al. Excitation of 2+ isomeric level of $^{104}$Ag nucleus in photonuclear reactions // *Proceedings of the International Conference on Nuclear data for Science and Technology,* EDP Sciences, Nice, France, 2008, p. (April 22-27, 2007).
64. D. Kolev, E. Dobreva, N. Nenov, V. Todorov. A convenient method for experimental determination of yields and isomeric ratios in photonuclear reactions measured by the activation technique // *Nucl. Instr. and Meth. in Phys. Res.* 1995, A356, p. 390, doi.org/10.1016/0168-9002(94)01319-5.
65. A.N. Vodin, O.S. Deiev, I.S. Timchenko, et al. Photoneutron cross-sections for the reactions $^{181}$Ta($\gamma$; xn; x=1…8)$^{181-x}$Ta at $\{E\}_{\gamma max}$=80-95 MeV // *Eur. Phys.*

*J.* 2021, v. A57, p. 208. arXiv:2103.09859, doi.org/10.1140/epja/s10050-021-00484-x.
66. V. Zheltonozhsky, M. Zheltonozhskaya, A. Savrasov, and A. Chernyaev. Book of Abstracts LXX Int. Conf. "NUCLEUS-2020" online part, 12-17 October 2020, Saint Petersburg, 64.
67. B.S. Ishkhanov, V.N. Orlin, and S.Yu. Troschiev // *Phys. Atom. Nucl.* 2012, 253, p. 75, doi.org/10.1134/S1063778812020093.
68. J.H. Carver and W. Turchinetz. The (γ,2n) and (γ,3n) reactions in $^{181}$Ta // Research School of Physical Sciences, Australian National University, Canberra, September 1957, p. 613.
69. H. Bartsch, K. Huber, U. Kneissl, and H. Krieger. Critical consideration of the statistical model analysis of photonuclear isomeric cross-section ratios // *Nucl. Phys.* 1976, A256, p. 243. doi.org/10.1016/0375-9474(76)90106-8.